# Investigating Quality of Institutional Repository Website Design Using Usability Testing Framework

Aang Subiyakto[1, a)], Yuliza Rahmi[1, b)], Nia Kumaladewi[1, c)], M. Qomarul Huda[1, d)], Nidaul Hasanati[1, e)], Tri Haryanto[2, f)]

[1]*UIN Syarif Hidayatullah Jakarta, Jl. Juanda 95, Tangerang Selatan, 15412, Indonesia*
[2]*Jasa Raharja Insurance,, JL Rasuna Said Kav. C-2, Jakarta Selatan, 12920, Indonesia*

[a)]aang_subiyakto@uinjkt.ac.id, [b)]yuliza.rahmi14@mhs.uinjkt.ac.id, [c)]nia.kumaladewi@uinjkt.ac.id, [d)]mqomarul@uinjkt.ac.id, [e)]nidaul.hasanati@uinjkt.ac.id, [f)]tri.haryanto@jasaraharja.co.id

***Abstract.*** *Quality of website design is one of the influential factors of website success. How the design helps the users using effectively and efficiently website and satisfied at the end of the use. However, it is a common tendency that websites are designed based on the developer's perspectives and lack considering user importance. Thus, the degree of website usability tends to be low according to user perceptions. This study purposed to understand the user experiences using an institutional repository (IR) website in a public university in Indonesia. The research was performed based on usability testing framework as the usability testing method. About 12 participants were purposely involved concerning their key informant characteristics. Following three empirical data collection techniques (i.e., query technique, formal experiment, and thinking aloud), both descriptive analysis using usability scale matric and content analysis using qualitative data analysis (QDA) Miner Lite software were used in the data analysis stage. Lastly, several visual design recommendations were then proposed at the end of the study. In terms of a case study, besides the practical recommendations which may contextually useful for the next website development; the clarity of the research design may also help scholars how to combine more than one usability testing technique within a multi-technique study design.*

## INTRODUCTION

In the human and computer interaction study, system usability is related to how a system can work well when used maximally by users so that all system capabilities can be useful as well. Thus, the user involvements may the essential factors in the assessment (1-3). Usability testing is a technique to get direct opinions from users by assigning them to complete different tasks on the real system and take feedback (4). The technique is useful in user-based systems. In general, a user-based system is a system that can generate profits because companies will need users of the systems they build for their business continuity, but that does not mean that systems that do not come from non-profit companies do not need to be concerned with user aspects. Educational institutions such as universities also need to pay attention to the user aspects because several systems can be accessed by the general public, especially their systems that can be opened through the website. With frequent websites college opened, it will be able to help increase the college's ranking, such as ranking methods conducted by the Webometrics (5). The visitor number of educational institution websites is one of the highlight variables here.

Concerning to the abovementioned phenomena, the researchers carried out a preliminary study by observing the visitor number of a public university website in Indonesia and found that around 53% of the visitors were the institutional repository (IR) website of the university. In the deep website traffic analysis; even though the website has a lot of visitor traffic but almost half of the total visitors (42%) did not return to access the website. Besides, almost 49% of visitors immediately left the IR website when they only accessed one page (bounce rate). This may be a quite high ratio for a website bounce. The high traffic bounce rate on the IR website can be caused by several reasons, including the lack of usability on the website. Making interface design changes that pay attention to aspects of usability can reduce bounce rates on the IR website. Furthermore, interviews with ten users of the IR indicated that the complaints around the IR interface are still dominated the use report, including search results that are not following what has been typed, the location of the button is confusing, the use of difficult filters, the layout is not neat, and the appearance is not attractive.

Therefore, the usability testing may indispensable to be done to explore the problems experienced by users when using the IR website and to propose the system interface modification. Here, complains and responses of the users who participated in the study were the objective criteria of the study. It was why usability testing framework used in this qualitative study. There are two questions which guided the research implementation:

RQ1: How is the level of quality of the IR website based on the usability testing framework used in the study?
RQ2: Considering to observations of the system used by its users, what are the proposed recommendations?

Sequentially, the next parts of this article describe the methodological points of the study in the second part, results and discussion in the third part which also elucidates recommendations based on the comparative interpretations with the other website, and conclusion part at the end of the paper.

## RESEARCH METHODS

This qualitative study was carried out based on usability testing research framework adopted from Isa, Lokman (6), Thatawong and Jiamsanguanwong (7), and Navarro, Molina (8). Besides the significant role of the IR in the higher education institution, the IR website of the university was selected based on the findings of the preliminary study step. Figure 1 shows the procedure of this study. The study consisted of six main steps, including the preliminary study step until the research report writing one by adopting the above-mentioned framework. Around 12 participants were involved purposively by considering their key informant role (9-11), regarding the activity level using the IR website.

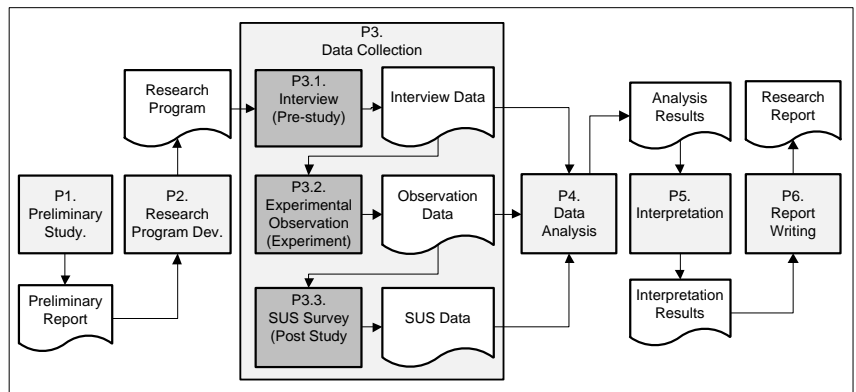

**FIGURE 1**. Research procedure

A preliminary study (P1) was carried out by studying similar literature and then observing the visitation traffic of the university website domain, to identify the unique visitor phenomenon. The researchers then developed research programs in terms of the research proposal for proposing the research implementation (P2). Here, the usability testing framework was then selected by the researchers.

Following the research framework; the three qualitative data collection steps were then implemented in this study (P3). In the pre-study (structured interview) steps (P3.1), the participants were directed to answer a structured interview around their data and general knowledge about the IR website. It was to capture demographic data of the participants, including personal data, internet usage data, and participant's general knowledge about the website (Table 4). In the experimental observation step (P3.2), the participants were directed to complete five task scenarios (Table 2), their activities were observed to measure the efficiency and effectiveness values of the performance, and their thinking out loud responses was noted during the task completion using the think-aloud technique). Camtasia Studio 8 was used in this stage to record the participant activities. In the last data collection step (P3.3), the participants were then given the system usability scale (SUS) Questionnaires (Figure 2.a) to assess the satisfaction value of the performance.

**Table 2.** The scenario of the experimental observation

| Scenario | Aims |
|---|---|
| 1. You are a researcher who wants to read research on "The effects of Islamic religious education." How can you find and read one of the studies on the topic on the IR website? | Knowing how users find a research topic you want to read. |
| 2. You are in the mood to read thesis research from the psychology majors as a reference for your next research topic. How can you find thesis researches from the department? | Knowing how users find specific types of research (thesis) |
| 3. You are interested in reading the results of research conducted by Azyumardi Azra. How can you find and read one of the studies by the author mentioned? | Knowing how users find specific types of research (thesis) |
| 4. Previously you have found research on "The effects of Islamic religious education." However, you still want to read research on the topic but by limiting the findings so that the title of the research is not related to the word "economy". How can you find and read it? | Find out how users use the add filters feature |
| 5. You are also interested in reading research on Juvenile Delinquency. However, you want to read the research based on the most recent research made. How can you find and read one of the studies on that topic? | Find out how users can find research by year. |

**Table 3.** SUS questionnaires

| Statements/Questions | 1 | 2 | 3 | 4 |
|---|---|---|---|---|
| 1. I think I will use this system again | O | O | O | O |
| 2. I feel this system is complicated to use | O | O | O | O |
| 3. I feel this system is easy to use | O | O | O | O |
| 4. I need help from other people or technicians for using this system | O | O | O | O |
| 5. I feel the features of this system are working properly | O | O | O | O |
| 6. I feel, many things are inconsistent (incompatible) in this system | O | O | O | O |
| 7. I feel other people will understand how to use this system quickly | O | O | O | O |
| 8. I find, this system confusing | O | O | O | O |
| 9. I feel, there are no obstacles for using this system | O | O | O | O |
| 10. I need to get used to first before using this system | O | O | O | O |

In the analysis step, the effectiveness value was calculated based on the completion rate of the given tasks. The completion rate was calculated by setting a binary value of '1' if the participant succeeds in completing the task and '0' if unsuccessful. The threshold value of the recommended task is 78%. The efficiency value was assessed based on the time taken by participants to complete each scenario. Meanwhile, the thinking out loud responses were analyzed using thematic analysis using the QDA Miner Lite software. Moreover, to calculate the SUS score, first, add up the contribution score of each item. For odd items, the contribution of the score is the scale position minus 1. For even items, the contribution is 5 minus the scale position. Multiply the total score by 2.5 to get the overall SUS score.

In the interpretation step, the researchers then interpreted the results based on the threshold values and compared the IR website design with the similar one (i.e., EEE, JSTOR, DOAJ, Scopus, and IR websites of two public universities in Indonesia), in order to propose recommendations. Here, the Axure RP 8 was used to design the recommendation points.

## RESULTS AND DISCUSSION

Table 4 presents the profiles of the 12 participants in this qualitative study. Most of the people were 17-25 years old bachelor degree, and from the internal institution (around 8 people, ±67%), daily using internet (about 11 persons, ±92%), and people who know and ever access (about 10 participants, ±84%).

## Performance Measurements

Efficiency was measured based on the task time required by participants to complete a task (Equation 1). Figure 5 shows the results of the time data required by participants to complete the task when doing usability testing (the time unit was second). The efficiency value in this assessment was 66%. The effectiveness calculation results (Equation 2) of the complete task shows a value of 76.66%. Table 5 and Table 6 show that T4 is the task that most participants cannot complete, followed by assignments T2, T5 and T1. Also, although in T1 only one participant failed to complete it, seven participants completed it with difficulty, as did T5. Whereas T3 is the only task that can be completed by all participants. Lastly, the satisfaction value of the system was assessed with SUS questionnaires. Figure 2.a. demonstrates the average value of the calculation in about 62.3 points.

**Table 4.** The scenario of the experimental observation

| Profiles | Characteristics | f | % | Profiles | Characteristics | f | % |
|---|---|---|---|---|---|---|---|
| Sex | Male | 6 | 50 | Institutions | UIN Jakarta | 8 | 66.7 |
|  | Women | 6 | 50 |  | Other Universities | 4 | 33.3 |
| Age | 17-25 years | 8 | 67 | Internet Use | Daily | 11 | 91.7 |
|  | 26-35 years | 1 | 8.3 |  | Three times a week | 1 | 8.3 |
|  | 36-45 years | 1 | 8.3 | Knowledge about the IR | Know and ever access | 10 | 83.3 |
|  | > 45 years | 2 | 17 |  | Know but never access | 2 | 16.7 |
| Status | Bachelor | 8 | 67 | Purpose of accessing the IR | Find the thesis | 5 | 41.7 |
|  | Postgraduate | 2 | 17 |  | Find journal | 5 | 41.7 |
|  | Lecturers | 2 | 17 |  | Find a dissertation | 1 | 8.3 |

**Table 5.** Execution Time of Completeness Task

| Task | P1 | P2 | P3 | P4 | P5 | P6 | P7 | P8 | P9 | P10 | P11 | P12 |
|---|---|---|---|---|---|---|---|---|---|---|---|---|
| T1 | 123 | 135 | 61 | 132 | 76 | 233 | 101 | 171 | 160 | 644 | 213 | 118 |
| T2 | 120 | 185 | 129 | 76 | 58 | 272 | 32 | 70 | 30 | 324 | 284 | 170 |
| T3 | 51 | 58 | 64 | 89 | 45 | 75 | 46 | 31 | 76 | 88 | 40 | 53 |
| T4 | 93 | 44 | 225 | 159 | 75 | 105 | 78 | 53 | 141 | 189 | 179 | 70 |
| T5 | 172 | 61 | 279 | 36 | 69 | 263 | 123 | 51 | 287 | 343 | 339 | 246 |

☐ : Success     ☐ : Failed

**Table 6.** Task Completeness

| Task | P1 | P2 | P3 | P4 | P5 | P6 | P7 | P8 | P9 | P10 | P11 | P12 |
|---|---|---|---|---|---|---|---|---|---|---|---|---|
| T1 | 1 | 1 | 1 | 1 | 1 | 1 | 1 | 1 | 1 | 0 | 1 | 1 |
| T2 | 1 | 1 | 0 | 1 | 1 | 1 | 1 | 1 | 1 | 0 | 0 | 0 |
| T3 | 1 | 1 | 1 | 1 | 1 | 1 | 1 | 1 | 1 | 1 | 1 | 1 |
| T4 | 1 | 1 | 0 | 1 | 0 | 0 | 0 | 1 | 1 | 0 | 1 | 0 |
| T5 | 1 | 1 | 1 | 1 | 0 | 1 | 0 | 1 | 0 | 1 | 1 | 1 |

1 : Success    2 : Failed
☐ : Success with easy    ☐ : Success with difficult    ☐ : Failed

$$\text{Overall Relatives Efficiency} = \frac{5490}{8313} \times 100\% = 66\% \quad \ldots \ldots \ldots \ldots \ldots \ldots \ldots \ldots \quad (1)$$

$$\text{Effectiveness} = \frac{46}{60} \times 100\% = 76,66\% \quad \ldots \ldots \ldots \ldots \ldots \ldots \ldots \ldots \ldots \ldots \quad (2)$$

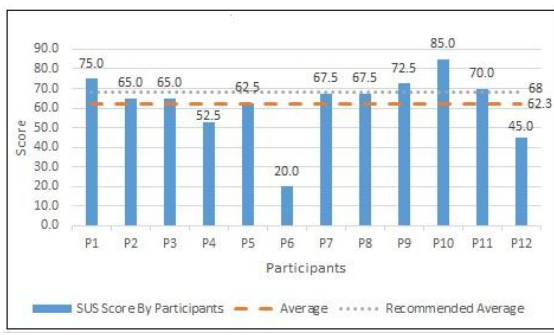
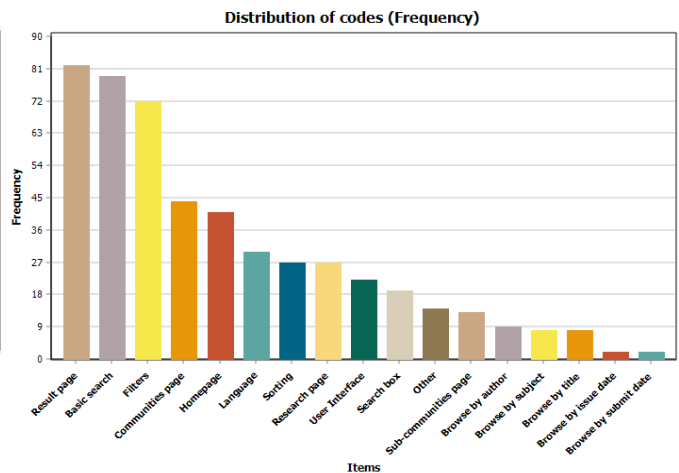

(a)            (b)
**FIGURE 2.** (a) SUS score (b) Distribution of codes (frequency)

In short, we can see that the result value of the overall relative efficiency calculation was 66%. It is higher than the recommended value of around 50% (12). Furthermore, it is different from the efficiency issue. The effectiveness value is still slightly below the average value which recommended Sauro and Lewis (13), that is 78%. Meanwhile, the satisfaction value (Figure 2.a) of the SUS score was 62.3 point. It is slightly below the score recommended by Brooke (14) which is 68 point. Moreover, in the qualitative data analysis from interviews, observations, and think aloud by using the miner lite tool, it was found that the five most frequently occurring topics were the result page, basic search, filters, communities, and homepage (Figure 2.b).

## Recommendations

### *Homepage*

Based on observations, most participants tended to be confused and then used basic search to complete the task which is given. In Figure 3.a., it can be seen that there is an advanced search feature on the recommended homepage. The justification was because the observation results and comments which indicated that it would be better if the feature is advanced search and filter placed at the beginning of the page without having to type keywords and do filtering on the result page.

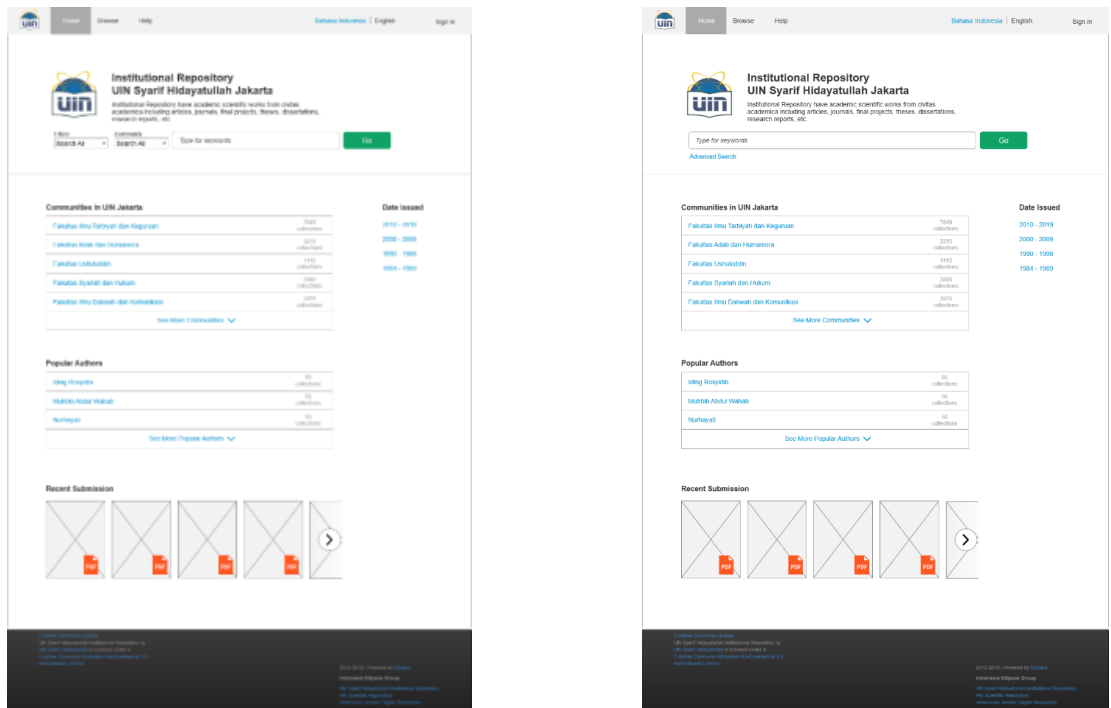

(a)        (b)

**FIGURE 3**. a) Recommendation Homepage   b) Recommendations filter and sorting

### *Basic Search*

The basic search feature is a feature that was often used by participants. Since the first time participants access the website, until when participants feel confused with the search results or feel they have not found what they are looking for, the participants will immediately type in the keyword in the basic search. Based on the results of the analysis, the recommendations shown in the user interface for basic search are given as Figure 3.b.

### *Filters, Sorting, and Search Boxes*

Features of filtering, sorting, and search boxes are the features that most often make participants confused and spend a long time when completing tasks. This is caused by the location button which is close to each other even though each button has a different feature, but many participants consider the button to have the same function. Based on the above-mentioned descriptions, Figure 3.a. presents the new design of the features.

*Communities Page*

In community pages, quite a lot of participants find it difficult because they feel confused with the many menu choices on the page so that in the end the participants prefer to use the feature basic search in completing the task the given. Besides, the layout on the page also looks neat. Based on the results of the analysis, the recommended user interface is to simplify the menus contained on the page and change the page layout (Figure 3.b.)

*Result Page*

Based on observations, many participants were confused and spent a lot of time on the result page. Some of the reasons are that participants are distracted by the number of suggestion links for author, community, and issue date. Figure 4.a. shows that link suggestion was removed to make the user focused on the results data.

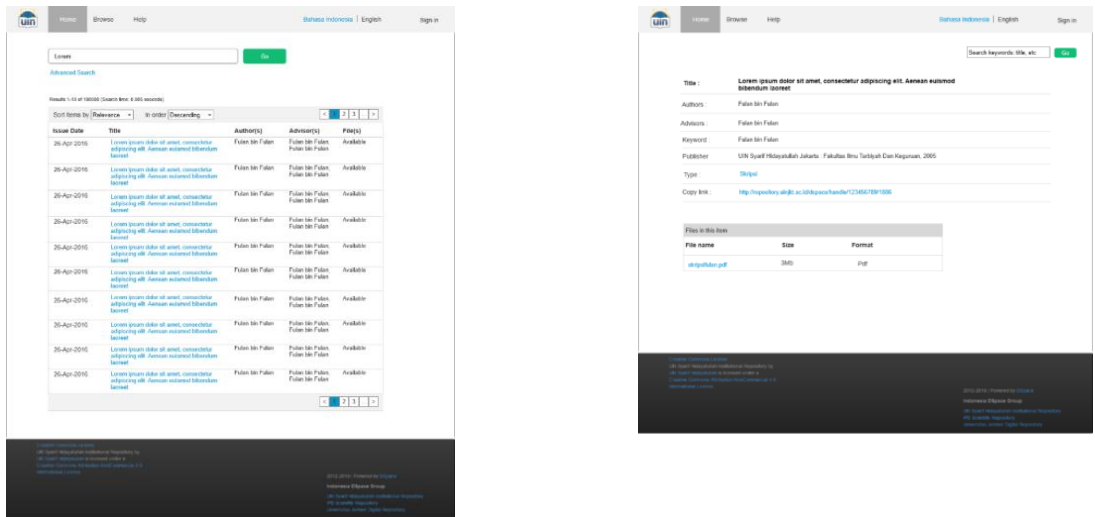

(a) (b)
**FIGURE 4**. a) Recommendations result from page   b) Recommendations research page

*Research Page*

Based on observations towards the research page, participants have problems in the identifier link section at the top because participants think that the link is a download link. Also, the URL section also often makes participants misunderstand because participants also think the URL is a download link. Then the button below is also quite often made the participants mistaken that the button is a button to download especially when the file is not available. Thus, it was recommended to remove the section link identifier and URL to simply replace it with a link (Figure 4.a.)

## CONCLUSION

In summary, the value of the overall relative efficiency calculation was 66%. It is higher than the efficiency threshold value. Meanwhile, the effectiveness value is still slightly below the threshold value used in the usability studies in about 78%. Similar to the effectiveness point, the satisfaction value (around 62.3 points) was also slightly below the score recommended value in about 68 points. The above-mentioned results of the usability testing framework may have answered the first question of the study. Despite the IR website was efficient, but it was ineffective and unsatisfied in the implementation. It may also have related to the results of the observation (think aloud) data analysis. Referring to the second research question, the qualitative analysis results indicated five most frequently occurred topics, i.e., the result page, basic search, filters, communities, and homepage. Therefore, the researchers then proposed the six recommendations around the topic based on the comparative interpretation with similar websites. It may be a practical contribution to the study. Of course, besides the employments of the methodological points and the researcher's competencies, the involvements of the 12 participants may also have the limitations here. Therefore, they may be the consideration points for the next studies.